\documentstyle[12pt]{article}

\begin{document}
\begin{titlepage}
\title{The orthogonality relations for the supergroup  $U(m|n)$}
\author{Jorge Alfaro \thanks{e-mail address:
jalfaro@lascar.puc.cl .} \ and Ricardo Medina \\
Facultad de F\'\i sica\\Universidad Cat\'olica de
Chile\\Casilla 306, Santiago 22, Chile
\\
and\\
Luis F. Urrutia \thanks{On leave of absence from Instituto de
Ciencias
Nucleares, UNAM, Mexico.} \\
Departamento de F\'\i sica\\
Universidad Aut\'onoma Metropolitana-I\\
Apartado Postal 55-534\\
09340 M\'exico, D.F.\\
and \\
Centro de Estudios Cient\'\i ficos de Santiago \\
Casilla 16443, Santiago 9, Chile}
\maketitle
\begin{abstract}
Starting from the generalization of the Itzykson-Zuber integral
for  $U(m|n)$ we determine the orthogonality
relations for this supergroup.
\end{abstract}
\vfill
\begin{flushleft}
SHORT TITLE: Orthogonality relations for $U(m|n)$. \\
PUC-FIS 35,\ May 1995.
\end{flushleft}
\end{titlepage}
\newpage

\baselineskip=20pt

Motivated by the recent progress made in the study of
random surfaces and statistical systems on random surfaces,which might
have important applications in non-critical string theory as well as
Quantum Chromodynamics (QCD) in the large $N$ limit, we have considered
the extension of some of these ideas to the case where the associated
random matrices \cite{mehta1} are replaced by
supermatrices.
An important mathematical object that appears naturally in the discussion of
random matrices is the Itzykson-Zuber (IZ)
 integral over the unitary group \cite{IZ}. This integral
has been applied to the solution of the two-matrix model \cite{IZ},
\cite{mehta2}
and, more recently, to the Migdal-Kazakov model of "induced QCD"
\cite{migkaz}.

 Recently we have extended the IZ  integral to the case of the
unitary supergroup $U(m|n)$ \cite{AMU}. In this letter we apply this
result to determine the orthogonality relations among  irreducible
representations of this supergroup. The basic problem that arises is that
the integration measure $[dU]$ over $U(m|n)$ is of the Berezin type,
which include
integrations over odd Grassmann numbers according to the standard recipe
\cite{DEWITT}. Thus, in many cases the integration over $[dU]$ of
supermatrix elements corresponding to arbitrary representations of the
supergroup will be automatically zero due to the above mentioned
Grassmannian character.
In particular, this will happen in the case of the orthogonality relations
and the purpose of this letter is to characterize the irreducible
representations of $U(m|n)$ which lead to a non-zero result together with
the determination of the  corresponding normalization coefficient.

In the following paragraphs we briefly summarize
our conventions regarding
representations of supergroups together with some results that
will be used subsequently.

Supergroups will be represented by linear operators ${\tilde D}(g)$ acting
on some vector space with basis $ \{ \Phi_I \}$. Linearity is defined
by ${\tilde D}(g) \left(\Phi_I \alpha  +\Phi_J \beta   \right) =
\left( {\tilde D}(g)\Phi_I\right)\alpha + \left( {\tilde
D}(g)\Phi_J\right)\beta$,
where $\alpha$ and $\beta$ are arbitrary Grassmann numbers. The action
\begin{equation}
{\tilde D}(g)\left(\Phi_I \right)=\sum_{J}^{m_t +n_t} \Phi_J
{\cal D}^{(t)}_{JI}(g),\label{REP}
\end{equation}
defines a representation $\{t\}$ of the supergroup characterized by the
Young tableau $ (t_1, t_2, \dots  )$, with
$t_1 \geq t_2 \geq, \dots $, in the usual notation.
Here ${\cal D}^{(t)}_{JI}(g)$ are the elements
of an $(m_t +n_t)\times(m_t +n_t)$
supermatrix written in the standard block form \cite{DEWITT}.
In fact, our definition of linearity given above guarantees
that the definition
(\ref{REP}) satisfies ${\cal D}^{(t)}_{JI}(g_1*g_2)=\sum_K
{\cal D}^{(t)}_{JK}(g_1)
{\cal D}^{(t)}_{KI}(g_2)$, thus providing a representation
 of the supergroup
in terms of the standard multiplication of supermatrices.

The Schur's lemmas can be directly proved in  the case of finite
supergroups and the extension to continuous supergroups is made in
complete analogy to the classical case. In general, the corresponding
measure must be left and right-invariant under the supergroup action
and for the case of $U(m|n)$ it is defined by
$[dU] = \mu \prod_{P,Q=1}^{m+n} dU_{PQ} dU_{PQ}^* \
 \delta (UU^\dagger - I)$, where
the $\delta$-function really means the product of $(m+n)^2$ unidimensional
$\delta$-functions corresponding to the independent constraints set by the
condition $UU^\dagger = I$. The normalization constant
$\mu$ is fixed by our normalization of the supersymmetric IZ integral.
It is important to observe
that the above  measure possesses $2mn$ real independent odd
differentials.

The application of Schur's lemmas to the quantity
${\cal X}^{st}_{IJ}=\int [dU]{\cal D}^{(s)}_{IL}(U)\break X_{LM}
{\cal D}^{(t)}_{MJ}(U^{-1})$, where $X_{LM}$ is an arbitrary
 supermatrix, leads
to the conclusion that ${\cal X}^{st}_{IJ} $ must be a multiple
of the identity supermatrix. Factoring out the arbitrary piece
$X_{LM}$, we are left
with the ortogonality relations
\begin{eqnarray}
\int [dU] {\cal D}_{IJ}^{(s)} (U) {\cal D}_{KL}^{(t)*} (U) = (-1)^
{\epsilon_J} \alpha_{\{{t}\}} \delta^{st} \delta_{IK} \delta_{JL},
\label{ORTO}
\end{eqnarray}
where $(U^\dagger)_{ij}=(U^{-1})_{ij}=(U^{*})_{ji}$. Let us recall that there
are two fundamental representations of $U(m|n)$ : ${\cal D}_{ij}(U)=U_{ij}$
and ${\dot{\cal D}}_{ij}(U)={\tilde U}_{ij}=(-1)^{\epsilon_j+\epsilon_j
\epsilon_i}(U^*)_{ij}$. These lead to three types of irreducible
representations  $\{t\}$:  undotted  $\{u\}$, dotted  $\{{\dot v}\}$ and
mixed  $\{{\dot v}\}|\{u\}$
\cite{BARS}.

As compared to the classical case, the appearence of odd integration
variables in $[dU]$ imposes further contraints upon the representations
that give a non-zero value for the coefficient $\alpha_{\{{t}\}}$ in
(\ref{ORTO}). The main result of this letter is to characterize
such representations in the dotted and undotted cases.

To begin with, we prove that they satisfy the following
Lemma: The supercharacters $s\chi_{\{t\}} (U)\equiv \sum_I
(-1)^{\epsilon_I} {\cal D}_{II}^{(t)} (U) $
 of  the representations ${\cal D}_{IJ}^{(t)} (U)$
for which $\alpha_{\{ t \}}
\neq 0$ constitute a linearly independent set.
The proof goes as follows: the ortogonality relations
(\ref{ORTO}) imply that
\begin{equation}
\int[dU] s\chi_{\{s\}} (U) {\cal D}_{KL}^{(t)*}
(U) = \alpha_{t} \delta^{st} \delta_{KL}.
\label{ORTOC}
\end{equation}
Next, let us consider a null linear combination of supercharacters of
 representations with $\alpha_{\{s\}} \neq 0$:
$\sum_{s} a_{s} s\chi_{\{s\}} (U) = 0$.
Multiplying this equation by ${\cal D}_{kl}^{(t)*} (U)$ and integrating
over dU we have
$a_{t} \alpha_{\{{t} \}} \delta_{kl} = 0$ for each representation $\{t\}$,
which shows that $a_{t} = 0$ provided $\alpha_{\{{t} \}}\neq 0$ .

The starting point that leads to the determination of
 the undotted  representations $\{t\}$ together with  the values
of the non-zero $\alpha_{\{{t} \}}$ in (\ref{ORTO}) is our
supersymmetric extension
of the IZ integral given by \cite{AMU}
\begin{eqnarray}
\tilde{I} (M_1,M_2;\beta) \equiv \int [dU]
e^{\beta Str(M_1 U M_2 U^\dagger)} \nonumber \\
= \beta^{mn} \Sigma(\lambda_1, \bar{\lambda}_1)
\Sigma(\lambda_2, \bar{\lambda}_2) I(\lambda_1, \lambda_2; \beta)
I(\bar{\lambda}_1, \bar{\lambda}_2; -\beta),\label{SUSYIZ}
\end{eqnarray}
where $I(d_1, d_2;\gamma)$ is the standard IZ integral \cite{IZ}
\begin{eqnarray}
I(d_1, d_2;\gamma) =\gamma^{- \frac {N(N-1)}{2}}  \prod_{p=1}^{m-1} p!
\frac {det(e^{ \gamma d_{1i} d_{2j} })}{\Delta (d_1) \Delta (d_2)},\label{IZN}
\end{eqnarray}
and
\begin{eqnarray}
\Delta (d) = \prod_{i>j} (d_i - d_j)
,\qquad \Sigma (\lambda, \bar{\lambda}) =
\prod_{i=1}^m \prod_{\alpha =1}^n
(\lambda_i - \bar{\lambda}_{\alpha}).
\end{eqnarray}
Here $M_1$ and $M_2$ are $(m + n) \times (m + n)$
 hermitian supermatrices which can be diagonalized \cite{JAP}
and $\beta$ is a complex parameter. Our notation is such that the first $m$
eigenvalues of $M$
are identified by $\lambda_i$, while the remaining $n$ eigenvalues are
denoted
by ${\bar\lambda}_{\alpha}$. Such partition is characterized by the
following
parity assignment of the eigenvector components $V_P,\bar V_P:
\epsilon(V_P)= \epsilon(P),\epsilon(\bar V_P)=\epsilon(P)+1$.

A convenient way  of rewriting the standard IZ integral is in terms of its
expansion in characters of the corresponding
irreducible representations of the  group $U(m)$, \cite{IZ}
\begin{eqnarray}
I(\lambda_1, \lambda_2;\beta) = \sum_{\{n\}} \frac{\beta^{|n|}}{|n|!}
\frac{\sigma_{\{n\}}}{d_{\{n\}}} \chi_{\{n\}} (\lambda_1) \chi_{\{n\}}
(\lambda_2).
\label{char-IZ}
\end{eqnarray}
Following analogous steps,
we obtain the supercharacter expansion of
the expresion (\ref{SUSYIZ})
\begin{eqnarray}
\tilde{I}(M_1, M_2;\beta) = \sum_{\{ t \} } \frac{\beta^{|t|}}{|t|!}
\sigma_{\{ t \} } \alpha_{\{t\}} s\chi_{\{t\}}(M_1) s\chi_{\{t\}}(M_2),
\label{char-SUSY-IZ}
\end{eqnarray}
which contains only undotted representations.
The above relation has been obtained without the use of a completeness
relation for the supercharacters.
Here $|t|$ denotes the total
number of boxes in the Young tableau corresponding to the
irreducible representation ${\{ t \} }$ of $U(m|n)$ and
$\sigma_{\{ t \}}$ counts the number of times that this representation
is contained in the tensor product ${\otimes}^{|t|}{\cal D}$.

In virtue of the Lemma previously proved, we see that the representations
which contribute to Eq.(\ref{char-SUSY-IZ}) have supercharacters that
form a linearly independent set.

Now we consider the determination of the representations
with non-zero $\alpha_{\{t\}}$.
The basic expression we use is the character expansion in both sides
of Eq.(\ref{SUSYIZ}), which is
\begin{eqnarray}
\sum_{\{ t\}} \frac{\beta^{|t|}}{|t|!} \sigma_{\{t\}} \alpha_{\{t\}}
s\chi_{\{t\}}(M_1) s\chi_{\{t\}}(M_2) = \sum_{\{p\}} \sum_{\{q\}}
\frac{\beta^{|p|+|q|+mn}}{|p|! |q|!} \frac{\sigma_{\{p\}} \sigma_{\{q\}} }
{d_{\{p\}} d_{\{q\}}} (-1)^{|q|} \times \nonumber \\
\times \Sigma(\lambda_1,\bar{\lambda}_1) \chi_{\{p\}} (\lambda_1)
\chi_{\{q\}} (\bar{\lambda}_1) \Sigma(\lambda_2, \bar{\lambda}_2)
\chi_{\{p\}} (\lambda_2) \chi_{\{q\}} (\bar{\lambda}_2).
\label{main}
\end{eqnarray}
Now we analize this equation by considering the following cases:
\subsubsection*{a) Case of $|t|< mn$}
Before making any further analysis, from (\ref{main}) we can immediately
conclude that
\begin{equation}
\alpha_{\{t\}} = 0, \ \mbox{for} \ |t|= 0,1,\ldots ,(mn-1).\label{TMENOR}
\end{equation}
because in both sides of this equation we have a power series in $\beta$,
and the right term of it starts with $\beta^{mn}$ while the left one
starts with $\beta^0$. The proof goes by assuming that some coefficients
$\alpha_{\{t\}}$ are non-zero.  The linear independence of them, together with
the above observation, imply that they must be zero, thus leading to a
contradiction.

\subsubsection*{b) Case of $|t| \geq mn$}
As we said just before, Eq.(\ref{main}) is  a power series in
$\beta$ in both sides of the equation, so for a same power of $\beta$ we must
have the same coefficient
\begin{eqnarray}
\frac{1}{|t|!} \sum_{\{t\}} \sigma_{\{t\}} \alpha_{\{t\}} s\chi_{\{t\}}(M_1)
s\chi_{\{t\}}(M_2) = \sum_{\{p\}} \sum_{\{q\}} \frac{(-1)^{|q|}}{|p|! |q|!}
\frac{\sigma_{\{p\}} \sigma_{\{q\}} }{d_{\{p\}} d_{\{q\}}} \times \ \ \
\ \ \ \ \ \ \nonumber \\
\times \Sigma(\lambda_1,\bar{\lambda}_1) \chi_{\{p\}} (\lambda_1)
\chi_{\{q\}} (\bar{\lambda}_1) \ \Sigma(\lambda_2, \bar{\lambda}_2)
\chi_{\{p\}} (\lambda_2) \chi_{\{q\}} (\bar{\lambda}_2),
\label{power}
\end{eqnarray}
where the sum in the LHS is made for all tableaux having a fixed
number of boxes $|t|$, while  the sum over $\{p\}$ and $\{q\}$ in the
RHS  is restricted to
\begin{equation}
|p|+|q| = |t| - mn.
\label{restriction}
\end{equation}
We now want to prove that Eq.(\ref{power}) necessarily implies that
\begin{eqnarray*}
s\chi_{\{t\}}(M) = c_{\{p,q\}} \Sigma(\lambda, \bar{\lambda})
\chi_{\{p\}} (\lambda) \chi_{\{q\}} (\bar{\lambda})
\end{eqnarray*}
for some $\{p\}$ and $\{q\}$ satisfying (\ref{restriction})
and for a certain representation  $\{t\}$ that we will determine.

In order to extract more information from Eq.(\ref{power}) let us consider
an arbitrary supermatrix $M_2$, while we restrict the supermatrix
$M_1$ in such a way that
one of its $\lambda$-eigenvalues  be equal
to one of its $\bar{\lambda}$- eigenvalues.
Namely, let $\lambda_j=\bar{\lambda}_{\beta}$, for example.
Then, in Eq.(\ref{power}) we are left with
\begin{eqnarray}
\frac{1}{|t|!} \sum_{\{t\}} \sigma_{\{t\}} \alpha_{\{t\}}
s\chi_{\{t\}}({M_1}) s\chi_{\{t\}}(M_2) = 0,
\label{null}
\end{eqnarray}
because $\Sigma(\lambda_1,\bar{\lambda}_1)$ becomes zero.
If we look at this relation as a null linear combination of the
supercharacters $s\chi_{\{t\}}(M_2)$ with coefficients
\begin{eqnarray}
\gamma_{\{t\}} = \frac{1}{|t|!} \sigma_{\{t\}} \alpha_{\{t\}}
s\chi_{\{t\}}({M_1}),
\label{gamma}
\end{eqnarray}
we  conclude that the coefficients  $\gamma_{\{t\}}$ are all
zero,
because the supercharacters appearing in (\ref{null}) constitute
a linearly independent set.
But $\sigma_{\{t\}}$ and $\alpha_{\{t\}}$ are different from zero, so that
we are left with $s\chi_{\{t\}}(\tilde{M_1})=0$.
Recalling that $s\chi_{\{t\}}(M)$ is a polynomial function of the
eigenvalues
$\lambda_i$,  $\bar{\lambda}_{\alpha}$,
we conclude from this relation that $s\chi_{\{t\}}(M)$ must be  divisible by
$(\lambda_j - \bar{\lambda}_{\beta})$. That is to say
\begin{eqnarray}
s\chi_{\{t\}}(M) = (\lambda_j - \bar{\lambda}_{\beta})
F_{j \beta}(\lambda,\bar{\lambda}),
\label{F}
\end{eqnarray}
where $F_{j \beta}(\lambda,\bar{\lambda})$ is another polynomial
function of the eigenvalues.
The same reasoning
can be extended to every $\lambda_i$ (i=1,...,m) and
$\bar{\lambda}_{\alpha}$ ($\alpha$ = 1,...,n), and this implies
that $s\chi_{\{t\}}(M)$
must have  the form
\begin{eqnarray}
s\chi_{\{t\}}(M) = \prod_{i=1}^m \prod_{\alpha =1}^n
(\lambda_i - \bar{\lambda}_{\alpha}) \ P(\lambda, \bar{\lambda})=
\Sigma(\lambda, \bar{\lambda})
\ P(\lambda, \bar{\lambda}).
\label{Sigma}
\end{eqnarray}
In Eq.(\ref{Sigma}), $P(\lambda,\bar{\lambda})$ must be  a homogeneous
polynomial function of all the eigenvalues,
because $s\chi_{\{t\}}(M)$ and $\Sigma(\lambda, \bar{\lambda})$ are so.
The degree of homogeneity of $s\chi_{\{t\}}(M)$ and $\Sigma(\lambda,
\bar{\lambda})$ is $|t|$ and $mn$, respectively. This means that
the degree of
homogeneity of $P(\lambda,\bar{\lambda})$ must be $|t|-mn$.
Also, we know that $s\chi_{\{t\}}(M)$ and $\Sigma(\lambda, \bar{\lambda})$
are symmetric functions in the eigenvalues $\lambda_i$,
$\bar{\lambda}_{\alpha}$, separately, and so should be
$P(\lambda,\bar{\lambda})$. Summing up then, $P(\lambda,\bar{\lambda})$ is:
(i) an homogeneous polynomial function of degree $|t|-mn$ in all the
eigenvalues and (ii)
a symmetric function of  $\{\lambda_i\}$ and
$\{\bar{\lambda}_{\alpha}\}$, separately.
Since the characters $\chi_{\{a\}}(\lambda)$ \
$\left(\chi_{\{b\}}(\bar{\lambda}) \right)$ are polynomial homogeneous
functions of degree $|a|$ \ $(|b|)$,
which are symmetric in the eigenvalues $\lambda_i$ \ $(\lambda_\alpha)$
and constitute a complete
linearly  independent set, $P(\lambda,\bar{\lambda})$
can be written as
\begin{eqnarray}
P(\lambda,\bar{\lambda}) = \sum_{\{a\},\{b\}} c_{\{a,b\}}^{\{t\}}
\chi_{\{a\}}(\lambda) \chi_{\{b\}} (\bar{\lambda}),
\label{P}
\end{eqnarray}
where the sum in $\{a\}$ and $\{b\}$ is rectricted by $
|a|+|b|=|t|-mn$.
Substituing this last relation in (\ref{Sigma}) we have
\begin{eqnarray}
s\chi_{\{t\}} (M) = \Sigma (\lambda, \bar{\lambda})
\sum_{\{a\},\{b\}} c_{\{a,b\}}^{\{t\}} \chi_{\{a\}}(\lambda)
\chi_{\{b\}} (\bar{\lambda}).
\label{new1}
\end{eqnarray}
Using the above expression in the LHS of
(\ref{power}) and comparing both sides of
this equation, we obtain  that the expansion in
(\ref{new1}) must include only one coefficient, for a given tableaux
${\{t\}}$, which precise form is yet to be determined.
That is
\begin{eqnarray}
s\chi_{\{t\}} (M) = c^{{\{t\}}}_{\{p,q\}}
\Sigma (\lambda, \bar{\lambda})
\chi_{\{p\}}(\lambda) \chi_{\{q\}} (\bar{\lambda}),
\label{new2}
\end{eqnarray}
where $\{p\}$ and $\{q\}$ satisfy (\ref{restriction}).
As the number
of solutions to Eq.(\ref{restriction})  is $|t|-mn+1$,
the supercharacter expansion of the
supersymmetric IZ integral will contain only ( $|t|-mn+1$ ) terms, for a
given $|t|$.
\subsubsection*{b.1) The case of  $\{p\} =
\{q\} = 0$ }
Here we have $|t|=mn$ and
\begin{equation}
s\chi_{\{t\}} (M) = c^{ {\{t\}}}_{\{0,0\}}
\Sigma (\lambda, \bar{\lambda}).
\label{c00}
\end{equation}
In order to proceed with the required identifications,
let us consider the particular case where the only non-zero block
of the supermatrix $M$ is the $m\times m$ block, i.e.
\begin{eqnarray}
M = \left ( \begin{array}{cc}
            M' & 0 \\
            0  & 0
            \end{array}
    \right ).
\label{case1}
\end{eqnarray}Then Eq.
(\ref{c00}) reduces to
\begin{eqnarray}
\chi_{\{t\}} (M') = c^{{\{t\}}}_{\{0,0\}}
(\prod_{i=1}^m \lambda_i )^n.
\label{c00M'}
\end{eqnarray}
Using Weyl's formula for the character of the representations of the
unitary group
\begin{eqnarray}
\chi_{\{r\}} (\lambda) = \frac {det (\lambda_i{}^{r_j +n -j})}
{det (\lambda_i{}^{n-j})}
\label{weyl}
\end{eqnarray}
we conclude that the product of eigenvalues in (\ref{c00M'})
corresponds to the character of the representation
 $\{r\}=(r_1,r_2,\ldots,r_m)$ with $r_i=n$
of $U(m)$. We are using the standard notation
$(r_1,r_2,\ldots,r_m)$ to denote a Young tableau with $m$ rows, such
that the i-th row has $r_i$ boxes.
In this way we have that
$\chi_{\{t\}} (M') = c_{\{0,0\}} \chi_{(n,n, \ldots ,n)} (M')$,
which allows the identification of the representation
$\{ t \}$ as the one given by the tableau corresponding to
 $ t_1=t_2= \ldots = t_m = n$, together with
$c^{\{t\}}_{\{0,0\}} = 1$.
Besides, we identify $\Sigma (\lambda, \bar{\lambda})$ as the
supercharacter of the representation referred to above. We will denote
by $\{mn\}$ the representation just found, whose tableau
consists of $m$ rows, each
with $n$ boxes.
\subsubsection*{b.2) The case $\{p\} \neq 0$,  $\{q\}=0$}
Here we have $|t|=|p|+mn$ and  $s\chi_{\{t\}} (M)=
c^{\{ t \}}_{\{p,0\}} \Sigma (\lambda, \bar{\lambda})
\chi_{\{p\}} (\lambda)$.
Considering in this expression the same choice of $M$ as in (\ref{case1}),
we have
$\chi_{\{t\}} (M') = c^{\{t\}}_{\{p,0\}} (\prod_{i=1}^m \lambda_i )^n
\chi_{\{p\}} (\lambda)$. Using again Weyl's formula we are able make the
identification $(\prod_{i=1}^m \lambda_i )^n \chi_{\{p\}} (\lambda) =
\chi_{\{n+p\}} (\lambda)$,
where by ${\{n+p\}}$ we mean the representation with Young
tableau $ (n+p_1, n+p_2,
\ldots , n+p_m )$. This leads to
$\chi_{\{t\}} (M') = c_{\{p,0\}} \chi_{(n+p_1, n+p_2, \ldots ,
n+p_m )} (\lambda)$ for this case and we  conclude that
$c^{\{t\}}_{\{p,0\}} = 1$ with ${\{t\}}$ being the representation
$( n+p_1, n+p_2, \dots , n+p_m )$ of $U(m|n)$. We introduce the pictorial
notation $\{n+p\}=\{mn\}\{p\}$, that will be useful in the sequel.
\subsubsection*{b.3) The case of arbitrary $\{p\}$ and $\{q\}$}
Now we discuss the main result of this letter which states that the
representations of $U(m|n)$ with $\alpha_{\{t\}}\neq0$  are characterized
by the
following Young tableaux
\begin{eqnarray}
\label{YT}
\{{\tilde t}\}=
              \begin{array}{c}
              \{mn\}\{p\} \\ \{q\}^T \ \ \ \ \
              \end{array}\equiv
\left({}^{\{p\}}_{\{q\}^T} \right),
\end{eqnarray}
with the normalization coefficient given by
\begin{eqnarray}
\alpha_{\{{\tilde t}\}} =
(-1)^{|q|} \frac{ |{\tilde t}|!}{|p|! |q|!} \frac{\sigma_{\{p\}}
\sigma_{\{q\}} }{\sigma_{\{{\tilde t}\}} } \frac{1}{d_{\{p\}} d_{\{q\}}}.
\label{finalresult}
\end{eqnarray}
The Young tableau $\{{\tilde u}\}=\left({}^{\{r\}}_{\{s\}^T} \right) $
introduced  in (\ref{YT})  is
constructed by starting from the basic array $\{n+r\}=\{mn\}\{r\}$
defined previously,
together with the array $\{s\}=\{s_1, s_2, \dots, s_n\}$,
which is subsequently transposed and
attached to the left bottom of it.

An important result
that leads to the above conclusions is that
\begin{eqnarray}
s\chi_{\{{\tilde t}\}} (M) = (-1)^{|q|} \Sigma (\lambda, \bar{\lambda})
\chi_{\{p\}}(\lambda) \chi_{\{q\}} (\bar{\lambda}).
\label{resultcpq}
\end{eqnarray}
Now we  give some details of the proof of Eq.(\ref{resultcpq}).
We start from the relation
\begin{equation}
s\chi_{\{mn\}\{u\}}(M)=\Sigma(\lambda,\bar{\lambda})
\chi_{\{u\}}(\lambda),\label{start}
\end{equation}
which is valid for every representationx $\{u\}$ of $U(m)$.
The proof will follow in two steps.
(i) First we prove, by induction, that
\begin{eqnarray}
s\chi_{\{{\tilde t}_1\}}(M) =(-1)^{|r_1|} \Sigma (\lambda, \bar{\lambda})
\chi_{\{p\}} (\lambda) \chi_{\{r_1\}^T }(\bar{\lambda}), \label{q}
\end{eqnarray}
where $\{{\tilde t}_1\}=\left({}^{\{p\}}_{\{r_1\}} \right)$
is a Young tableau of the type (\ref{YT})
with ${\{r_1\}}=(r_1)$, $r_1\leq n$,  corresponding to a single
row with  $r_1$
boxes, which is attached without transposition
to the bottom of $\{mn\}\{p\}$.
Let us consider  first the case $(1)=\{ \Box\}$, that is $ r_1=1$.
Taking $\{u\}=\{p\}$ in (\ref{start}) and multiplying both sides by
$s\chi_{\{\Box\}} (M) =
\chi_{\{\Box\}}(\lambda) - \chi_{\{\Box\}}(\bar{\lambda})$
we obtain
\begin{equation}
s\chi_{\{mn\}(\{p\} \times \Box )}(M) +
s\chi_{\{{\tilde t}_{11}\}}(M) = \Sigma (\lambda, \bar{\lambda})
\chi_{\{p\} \times \Box }(\lambda) - \Sigma (\lambda, \bar{\lambda})
\chi_{\{p\}} (\lambda) \chi_{\{\Box\}}(\bar{\lambda} ),\label{meanwhile}
\end{equation}
where $\{{\tilde t}_{11}\}=\left({}^{\{p\}}_{(1)} \right)$.
Applying (\ref{start})  to the case $\{u\}=\{p\} \times \Box  $
in (\ref{meanwhile}) we are left with
\begin{eqnarray}
s\chi_{\{{\tilde t}_{11}\}}(M)=- \Sigma (\lambda, \bar{\lambda})
\chi_{\{p\}}(\lambda) \chi_{\{\Box\}^T}(\bar{\lambda} ),
\end{eqnarray}
which verifies (\ref{q}) for this case. Here we  have made  use
of the Young tableux rules for multiplying representations.
Next  we asume that (\ref{q}) is  valid for the tableau
$ \{{\tilde t}_{1r}\}=\left({}^{\{p\}}_{(r)} \right)$,
and prove that it is also valid for
$ \{{\tilde t}_{1(r+1)}\}=\left({}^{\{p\}}_{(r+1)} \right)$
with $r+1 \leq n$.
For this purpose we will make use of the relation \cite{BARS}
\begin{eqnarray}
s\chi_{(n)}(M) = \sum_{k=0}^n (-1)^k \chi_{(n-k)}(\lambda)
\chi_{(k)^T }(\bar{\lambda}),
\label{schi}
\end{eqnarray}
where we recall that  $(k)$ denotes the Young tableau having
one row with $k$ boxes
while $(k)^T$ denotes de Young tableau corresponding to one column
with $k$ boxes. Considering this relation for $n=r+1$, separating the
$k=r+1$ term in the summation and multiplying
both sides of (\ref{start}) by (\ref{schi})we obtain
\begin{equation}
\sum_{k=0}^{r} s\chi_{\left({}^{\{p\}\times (r+1-k) }_
{(k)} \right)}(M) +
s\chi_{\left({}^{\{p\}}_{(r+1)} \right)}(M) = \sum_{k=0}^{r} (-1)^k \Sigma
(\lambda, \bar{\lambda}) \chi_{\{p\} \times (r+1-k)}(\lambda)$$
$$ \times \chi_{(k)^T}(\bar{\lambda})
+ (-1)^{r+1}  \Sigma (\lambda, \bar{\lambda})
\chi_{\{p\}}(\lambda) \chi_{(r+1)^T}(\bar{\lambda})
\end{equation}
The first terms of both sides are equal (in virtue of the hipothesis
of induction), so this last equation becomes the desired result.
(ii) Following analogous steps we can prove by induction in $r_2$, that
\begin{eqnarray}
s\chi_{\left({}^{{} \{p\}}_{(r_1,r_2)} \right)}(M)=(-1)^{r_1+r_2}
 \Sigma (\lambda, \bar{\lambda})
\chi_{{\{p\}}} (\lambda) \chi_{(r_1,r_2)^T}(\bar{\lambda})
\label{q1q2}
\end{eqnarray}
for $r_2 \leq n$.
The final choice $(r_1, r_2, \dots, r_n)^T=\{q\}$
implies the proof of the relation (\ref{resultcpq}), which after
substitution in (\ref{power}) leads to our final result (\ref{finalresult}).

An immediate consequence of our relation in (\ref{resultcpq})
is that we can obtain the dimension for the representations in
$U(m|n)$ that arise in the supercharacter expansion ($sd_{\{t\}}$)
in terms of the dimension of representations in $U(m)$ ($d_{\{p\}}$)
and $U(n)$ ($d_{\{q\}}$). Taking
\begin{eqnarray*}
M = \left( \begin{array}{rr}
              I_{m \times m}   &   0 \\
              0   &  -I_{n \times n}
             \end{array}
      \right)
\end{eqnarray*}
in (\ref{resultcpq}) and observing that
$s\chi_{\{t\}} (M)$  becomes $sd_{\{t\}}$, we obtain the closed expression
\begin{eqnarray}
sd_{\{{\tilde t}\}}= 2^{mn} d_{\{p\}} d_{\{q\}},
\label{superdimension}
\end{eqnarray}
for the dimensions of the representations of $U(m|n)$ characterized
by the tableaux in (\ref{YT}).
Let us also remark that our expression (\ref{finalresult}) correctly
reproduces the result $\alpha_{\{t\}}={1\over d_{\{t\}}}$ for $U(n)$. We end up
with a brief comment regarding the dotted and mixed representations. We are
able to prove that $\alpha_{\{\dot u \}}= \alpha_{\{u \}}$ and also that
$\alpha_{\{\dot u \}|{\{v \}}}$ can be written as a linear combination
of the undotted  coefficients $\alpha_{\{s \}}$. A detailed discussion of
these matters will be given in a
forthcoming publication.

\bigskip

LFU acknowledges the hospitality of J.Alfaro at Universidad Cat\'olica
de Chile. He is partially supported by the grants CONACyT(M\'exico)3544-
E9311 and UNAM-DGAPA-IN100694. RM acknowledges support from a CONICYT graduate
fellowship  and from the project FONDECYT 2950070.
JA acknowledges support from the projects FON\-DE\-CYT 1950809, Programa del
Gobierno Espa\~nol de Cooperaci\'on Cient\'\i fica con Iberoam\'erica and
a collaboration CNRS-CONICYT.

\end{document}